\documentstyle[eqsecnum,aps,preprint]{revtex}
\def\laq{\ \raise 0.4ex\hbox{$<$}\kern -0.8em\lower 0.62
ex\hbox{$\sim$}\ }
\def\gaq{\ \raise 0.4ex\hbox{$>$}\kern -0.7em\lower 0.62
ex\hbox{$\sim$}\ }

\def\NPB{{\em Nucl. Phys.} B}
\def\PLB{{\em Phys. Lett.}  B}
\def\PRL{{\em Phys. Rev. Lett. }}
\def\PRD{{\em Phys. Rev.} D}
\def\MPL{{\em Mod. Phys. Lett.}  A}
\begin{document}

\preprint{\vbox{\baselineskip=12pt
\rightline{BGU-PH-98/02}
\vskip0.2truecm
\rightline{CERN-TH/98-53}
\vskip0.2truecm
\rightline{hep-th/9803018}
\vskip1truecm}}

\vskip 2 cm

\title{Duality in Cosmological Perturbation Theory}

\author{R. Brustein${}^{1,2}$, M. Gasperini${}^{3,2}$ and
G. Veneziano${}^{2}$}
\vskip 2 cm
\address{(1) Department of Physics,
Ben-Gurion University,
Beer-Sheva 84105, Israel \\
(2) Theory Division, CERN, CH-1211 Geneva 23, Switzerland\\
(3)  Dipartimento di Fisica Teorica, Universit\`a di Torino,
 Via P. Giuria 1, 10125 Turin, Italy\\
 and Istituto Nazionale di Fisica Nucleare, Sezione di Torino, Turin,  
Italy\\
{\rm E-mail:} {\tt ramyb@bgumail.bgu.ac.il, gasperini@to.infn.it,
venezia@nxth04.cern.ch}}

\vskip 2 cm
\maketitle
\begin{abstract}
Cosmological perturbation equations derived
from low-energy effective actions are shown to be
invariant under a duality transformation reminiscent
of electric-magnetic, strong-weak coupling, S-duality.
A manifestly duality-invariant approximation for perturbations
far outside the horizon is introduced, and it is argued to be
useful even during a high curvature epoch.
Duality manifests itself through a remnant symmetry acting on the
classical moduli of cosmological models, and implying lower bounds on
the number and energy density of produced particles.

\end{abstract}


%%%%%%%%%%%%%%%%%%%%%%%%%%%%%%%%%%%%%%%%
\section{INTRODUCTION}
%%%%%%%%%%%%%%%%%%%%%%%%%%%%%%%%%%%%%%%%

In inflationary cosmology \cite{inflation}
 the Universe quickly becomes
homogeneous, isotropic, and spatially flat, but quantum  mechanics
 superimposes  on  the smooth classical backgrounds
calculable inhomogeneity perturbations  \cite{quantum}.
It is the purpose of this
 paper to point out a general duality symmetry obeyed by the  
lowest-order equations
 for these perturbations, and to show that duality may be used to
obtain lower bounds on the
energy density of the produced particles.

Our analysis applies to any physical perturbation $\Psi$
that is described, to lowest order
 in the strength of the perturbation and in its
derivatives, by the following quadratic action
\begin{equation}
I ={\small\frac{1}{2}} \int d\eta\ d^3x\ S(\eta) \left[ \Psi
^{\prime 2}- (\nabla \Psi )^2\right].
\label{spertact}
\end{equation}
In eq. (\ref{spertact}), which is generic in low-energy spatially flat
backgrounds, $\eta$ is the conformal time coordinate, and a prime
denotes $\partial/\partial\eta$. The function $S(\eta)$, sometimes
called the ``pump" field, is, for any given  $\Psi$, a known function of
the scale factor, $a(\eta)$, and of additional scalar fields (such as
a dilaton $\phi(\eta)$, moduli $b_i(\eta)$, etc.)
 on which the background may depend.

The duality symmetry, which, to the best of our knowledge, has not
been noticed before in the context of cosmological
perturbation theory, involves the interchange of the perturbation
$\Psi$ with its conjugate momentum $\Pi$, as well as the inversion of
the pump field $S$.  In some cases it
corresponds to a time-dependent version of $T$-duality, interchanging  
small and large
scale factors \cite{4,sfd}; in others it  corresponds to a
time-dependent version of
$S$-duality \cite{sdu,sdsc}, interchanging weak and strong coupling
and electric and
magnetic degrees of freedom; in most cases it amounts to a
time-dependent  mixture of both. As in other applications, duality turns
out to be a   powerful tool
for obtaining results that are hard to obtain otherwise.

Although most of our discussion will be completely general, it may be
useful to recall that, in the context of string cosmology,
 the massless perturbation
$\Psi$ may represent a physical polarization
of tensor metric perturbations (i.e. a gravity wave),
physical electromagnetic
perturbations  (i.e.  photons),  axionic perturbations, and so on.
In these three cases the pump field $S$ corresponds, respectively,
to the Einstein-frame scale factor \cite{1,7}, given in terms of the 
dilaton 
as $S=a^2 e^{-\phi}$, to the inverse string coupling \cite{2}
$S= e^{-\phi}$, and to the ``axion-frame" scale factor \cite{3} $S=a^2
e^{\phi}$. Here, and in what follows, $a$ is the scale factor of the
so-called string-frame metric, which coincides with the
usual scale factor of general relativity whenever the dilaton is
constant. For a generic string cosmology perturbation
 $S(\eta) =   a^{2m} e^{l\phi } b_i^{2m_i}$, where the numbers
$l, m, m_i$  depend  on the type of perturbation. Arguably,
the action (\ref{spertact}) provides a useful description for a much  
wider class of
perturbations.

%%%%%%%%%%%%%%%%%%%%%%%%%%%%%%%%%%%%%%%%
\section {Duality}
%%%%%%%%%%%%%%%%%%%%%%%%%%%%%%%%%%%%%%%%

In order to discuss the duality symmetry of
 the action (\ref{spertact}),
it is  convenient to use the Hamiltonian formalism.
The  Hamiltonian  corresponding to (\ref{spertact})
is given by
\begin{equation}
 H=\frac{1}{2} \int  d^3 x \Biggl[  S^{-1} \Pi^2 +
S (\nabla\Psi)^2\Biggr],
\label{ham}
\end{equation}
where $\Pi$ is the momentum conjugate to $\Psi$.
The first-order Hamilton equations  read
\begin{equation}
\Psi' = \frac{\delta H}{\delta \Pi}=  S^{-1} \Pi \; , ~~~~~~~~~~~
\Pi' = -\frac{\delta H}{\delta \Psi}= S \nabla^2\Psi \;,
\label{frstord}
\end{equation}
and lead to the decoupled second-order equations
\begin{equation}
\Psi''+\frac{S'}{S} \Psi'- \nabla^2\Psi = 0 \, , ~~~~~~~~~~~
\Pi''-\frac{S'}{S} \Pi'- \nabla^2\Pi = 0 \, ,
\label{psieq}
\end{equation}
of which only the first is commonly used in the study of
perturbations. In Fourier space the Hamiltonian (\ref{ham}) is given by
\begin{equation}
H=\frac{1}{2} \sum_{\vec{k}}
  \Biggl(  S^{-1} \Pi_{\vec{k}}\Pi_{-\vec{k}} +
S k^2 \Psi_{\vec{k}}\Psi_{-\vec{k}}\Biggr),
\label{fham}
\end{equation}
where $\Psi_{-\vec{k}} = \Psi_{\vec{k}}^*$ and $\Pi_{-\vec{k}} =
\Pi_{\vec{k}}^*$.
The equations of motion become
\begin{equation}
\Psi_{\vec{k}}' = S^{-1}{\Pi}_{-\vec{k}} \; , ~~~~~~~~~~~
\Pi_{\vec{k}}'= -S k^2 \Psi_{-\vec{k}} \; ,
\label{ffrstord}
\end{equation}
where $k=|\vec{k}|$. The transformation
\begin{equation}
\Pi_{\vec{k}} \rightarrow \widetilde \Pi_{\vec{k}}=k \Psi_{\vec{k}} \, ,
 ~~~~~ \Psi_{\vec{k}} \rightarrow \widetilde{\Psi}_{\vec{k}}=- k  
^{-1} \Pi_{\vec{k}}\, , ~~~~~
S \rightarrow \widetilde S=S^{-1},
\label{sduality}
\end{equation}
leaves the Hamiltonian, Poisson brackets,
and equations of motion unchanged.

We shall call this transformation, for short,
$S$-duality  since it contains the usual
strong-weak  coupling duality as a special case.  Indeed, by  
perturbing the tree-level, lowest-derivative string effective action,  
one finds
that, for  heterotic, four-dimensional gauge bosons, the   function $S$
is given simply by  \cite{2} $S(\eta)=e^{-\phi(\eta)}$. Since
$e^{\phi}$ determines the effective string coupling
constant, $e^{\phi}=g^2_{string}$, the transformation   $S \rightarrow
\widetilde S=S^{-1}$ amounts to performing, at each time $\eta$,  a
strong-weak coupling transformation $g_{string}\rightarrow
g_{string}^{-1}$, which is part of the $SL(2,Z)$  $S$-duality group
\cite{sdu}. Moreover, in the case of electromagnetic perturbations, the
contribution of $k\Psi_k$ to the Hamiltonian can be identified with that
of the magnetic field, whereas $\Pi$ is proportional to the electric
field,   hence, in
this case, the transformation (\ref{sduality}) exactly coincides with  
electric-magnetic duality.

We are interested in a situation where the time evolution of the pump
field amplifies the zero-point vacuum fluctuations of the field  
$\Psi$, satisfying the initial conditions
\begin{equation}
\langle S^{-1} \Pi^2\rangle= \langle S (\nabla\Psi)^2\rangle,
\label{incon}
\end{equation}
where $\langle\cdots\rangle$ denotes ensemble average, or
expectation value if perturbations are quantized. Under $S$-duality the
equations of motion are form-invariant, while the initial
conditions  are invariant  on the average in the sense of
eq. (\ref{incon}). Therefore,  under $S$-duality, solutions of the
perturbation equations
properly normalized to zero-point fluctuations,
are mapped into other normalized
solutions with the {\it same} total  Hamiltonian.
We will show that, as a consequence of this property, the energy
spectrum can be consistently estimated by truncating the solutions
outside the horizon to the frozen modes of $\Psi$ and $\Pi$.

This may be regarded as trivial, if one has in mind a
background where the non-frozen part of $\Psi$ quickly decays outside
the horizon, as in the standard inflationary scenario \cite{quantum}.
There are backgrounds, however, in which $\Psi$ grows in time instead
of decaying, and the growing mode of $\Psi$ dominates the
perturbation spectrum. Typical examples can be found in scalar-tensor
models of gravity and, in particular, in string cosmology \cite{6}. In
this context, the possibility of a consistent truncation to the frozen
part of the perturbations is a  powerful result. Only in a
duality-invariant approach is such a truncation
consistent, since, as shown below,
 the contribution to the Hamiltonian of the
growing mode of $\Psi$ is simply replaced by the  contribution
of the frozen part of its duality-related conjugate momentum.

%%%%%%%%%%%%%%%%%%%%%%%%%%%%%%%%%%%%%%%%
\section{Approximate solutions inside and outside the horizon}
%%%%%%%%%%%%%%%%%%%%%%%%%%%%%%%%%%%%%%%%

In order to solve the perturbation equations, and to normalize the
spectrum, it is convenient to introduce the canonical variables
${\widehat\Psi}$, ${\widehat\Pi}$,  whose  Fourier modes are  defined
by
\begin{equation}
\widehat\Psi_k = {S}^{1/2}\ \Psi_k \, , ~~~~~~~~~~~
\widehat\Pi_k = {S}^{-1/2}\ \Pi_k \; ,
\label{nv}
\end{equation}
so that the Hamiltonian density takes the canonical form
\begin{equation}
H=\frac{1}{2} \sum_{\vec{k}}
 \left( |{\widehat\Pi}_k|^2 + k^2 |{\widehat\Psi}_k|^2\right) .
\label{313}
\end{equation}
Under $S$-duality these canonical variables
 transform  as  the original variables. They
  satisfy  Schr\"odinger-like equations:
\begin{equation}
{\widehat\Psi}_k{''}+\left[k^2-(S^{1/2}){''}
S^{-1/2}\right]{\widehat\Psi}_k
= 0 \, ,  ~~~~~
{\widehat\Pi}_k{''}+ \left[k^2-(S^{-1/2}){''} S^{1/2}\right]
{\widehat\Pi}_k = 0.
\label{nscndord}
\end{equation}

The amplification of perturbations is typically associated with a
transition from an inflationary phase in which the pump field is
accelerated (in cosmic time), to a post-inflationary phase in which
the pump field is decelerated or constant. In such a class of
backgrounds, the ``effective potentials"
$V_{\Psi}=(S^{1/2}){''}S^{-1/2}$, $V_{\Pi}=(S^{-1/2}){''} S^{1/2}$,
grow  (in absolute value) during the phase of accelerated evolution, and
decrease in the post-inflationary, decelerated epoch, vanishing
asymptotically for very early times $\eta \rightarrow -\infty$
and vanishing again for
 very late times $\eta \rightarrow + \infty$.

The initial evolution of perturbations, for all modes with
$k^2> |V_{\Psi}|$, $|V_{\Pi}|$, may be described by the
WKB-like approximate solutions of eqs. (\ref{nscndord})
\begin{eqnarray}
{\widehat\Psi}_k(\eta)&=&\left( k^2-V_{\Psi}\right)^{-1/4}\
e^{\ -i\int\limits_{\eta_0}^{\eta} d\eta' \left( k^2-V_{\Psi}\right)^
{1/2} },  \nonumber \\
{\widehat\Pi}_k(\eta)&=& k \left( k^2-V_{\Pi}\right)^{-1/4}\
e^{\ -i\int\limits_{\eta_0}^{\eta} d\eta' \left( k^2-V_{\Pi}\right)
^{1/2} },
\label{wkbsol}
\end{eqnarray}
which we have normalized to a vacuum fluctuation, and
where the extra factor of $k$ in the solution for ${\widehat\Pi}_k$
comes  from consistency with the first-order equations (\ref{ffrstord}).
We  have ignored
a possible relative phase in the solutions. Solutions (\ref{wkbsol})  
manifestly
preserve the $S$-duality symmetry of the equations, since the
potentials  $V_{\Psi}$, $V_{\Pi}$ get interchanged under $S \rightarrow
S^{-1}$.

In the case of power-law evolution of the pump field, $S^{1/2} \sim
\eta^\alpha$, the effective potentials grow like $\eta^{-2}$ as $\eta
\rightarrow 0$, and the approximate solutions (\ref{wkbsol}) are valid,
at a given time $\eta$, for all modes $k$ with $(k\eta)^2 >1$.  These
modes are usually referred to as being ``inside the horizon". The
corrections to solutions (\ref{wkbsol}) are of order
${V_{\Psi}'}^2/{\left( k^2-V_{\Psi}\right)^{3}}$,
${V_{\Psi}{''}}/{\left( k^2-V_{\Psi}\right)^2}$ and higher, which are
indeed small for large $(k\eta)^2$. Asymptotically, at $\eta
\rightarrow -\infty$, we have $V_{\Psi}$, $V_{\Pi} \rightarrow 0$ and
solutions (\ref{wkbsol}) reduce to correctly normalized vacuum
fluctuations \begin{equation}
{\widehat\Psi}_k(\eta) =  k^{-1/2}\
e^{\ -i k\eta+ i\varphi_{\vec{k}} } \, , ~~~~~~~~~~
{\widehat\Pi}_k(\eta) =  k^{+1/2}\
e^{\ -i k\eta+ i\varphi_{\vec{k}}'},
\label{wkbsol1}
\end{equation}
where $\varphi_{\vec{k}}$, $\varphi_{\vec{k}}'$ are random phases,
originating from the random initial conditions. Note that because of the
random phases, $S$-duality  holds only on the average in the sense of eq.
(\ref{incon}).

For $k^2< |V_{\Psi}|$, $|V_{\Pi}|$, equivalently $(k\eta)^2<1$ if
$S^{1/2} \sim
\eta^\alpha$, the modes are usually referred to as being  ``outside  
the horizon".
For such modes,  it is possible to write
``exact" solutions to
eq. (\ref{nscndord}) as follows,
\begin{eqnarray}
{\widehat\Psi}_k(\eta)&=& \ S^{1/2} \ \ \Biggl[\widehat
A_k\ T\!\cos(S^{-1},S)+  \widehat B_k\
T\!\sin(S^{-1},S)\Biggr],\nonumber \\
{\widehat\Pi}_k(\eta)&=& k
S^{-1/2}\Biggl[\widehat B_k\ T\!\cos(S,S^{-1})-\widehat A_k\
T\!\sin(S,S^{-1})  \Biggr],
 \label{wkbsol3}
\end{eqnarray}
where $\widehat A_k, \widehat B_k$ are arbitrary integration
constants, and the functions
$T\!\cos(S^{-1},S; k, \eta)$,  $T\!\sin(S^{-1},S; k, \eta)$
are defined through the following expansions:
\begin{eqnarray}
&&T\!\cos(S^{-1},S; k, \eta)=
1- k \int\limits_{\eta_{ex}}^\eta d\eta_1 S^{-1}(\eta_1)
\ k \int\limits_{\eta_{ex}}^{\eta_1} d\eta_2 S (\eta_2)
 \nonumber \\
&+&
k \int\limits_{\eta_{ex}}^\eta d\eta_1 S^{-1}(\eta_1)
\ k \int\limits_{\eta_{ex}}^{\eta_1} d\eta_2 S (\eta_2)
\ k \int\limits_{\eta_{ex}}^{\eta_2} d\eta_3 S^{-1} (\eta_3)
\ k \int\limits_{\eta_{ex}}^{\eta_3} d\eta_4 S (\eta_4) +\cdots \;
;\nonumber \\
&&T\!\sin(S^{-1},S; k, \eta)=
k \int\limits_{\eta_{ex}}^\eta d\eta_1 S^{-1}(\eta_1)
- k \int\limits_{\eta_{ex}}^\eta d\eta_1 S^{-1}(\eta_1)
\ k \int\limits_{\eta_{ex}}^{\eta_1} d\eta_2 S (\eta_2)
 \ k \int\limits_{\eta_{ex}}^{\eta_2} d\eta_3 S^{-1} (\eta_3)
\nonumber \\
&+& \cdots .
\label{tsincos3}
\end{eqnarray}
The time $\eta_{ex}$, appearing as an arbitrary
lower limit of integration in (\ref{tsincos3}),
is most conveniently chosen to be near
horizon-exit time, i.e. such that $k\eta_{ex}\sim 1$.
The above functions satisfy
\begin{equation}
 \left[T\!\cos(S^{-1},S)\right]'= -\frac{k}{S} \ T\!\sin(S,S^{-1}) 
\, ,~~~~~
\left[T\!\sin(S^{-1},S)\right]'= \frac{k}{S} \ T\!\cos(S,S^{-1}).
\label{tsc}
\end{equation}
Using these relations,
and similar ones for $\left[T\!\cos(S,S^{-1})\right]'$ and
$\left[T\!\sin(S,S^{-1})\right]'$, it is possible to verify 
explicitly that 
(\ref{wkbsol3}) are indeed
solutions of eqs. (\ref{nscndord}). Formally, these solutions are also
 valid inside the horizon,  but the power series defining  $T\!\cos$,  
$T\!\sin$ are not
convergent there.
Solutions (\ref{wkbsol3}) manifestly preserve $S$-duality, which
now acts on the parameters of the solutions in the following simple
way:
\begin{equation}
\widehat A_k \rightarrow - \widehat B_k \; ,  ~~~~~~~~~~~~
\widehat B_k \rightarrow \widehat A_k \, ,  ~~~~~~~~~~~~
S \rightarrow S^{-1}\, .
\end{equation}

We now fix the integration constants by matching solutions
(\ref{wkbsol}) and (\ref{wkbsol3}) and their first derivatives at
horizon-exit time, for which $k\eta_{ex}\sim 1$. The result
(written for the  variables without the hats) is simply
\begin{eqnarray}
 {\Psi}_k(\eta)&=&  \left( k^2-V_{\Psi}^{ex}\right)^{-1/4} \ \Biggl[
S_{ex}^{-1/2} \ T\!\cos(S^{-1},S) e^{i\varphi_{\vec{k}} }+
S_{ex}^{1/2}\ T\!\sin(S^{-1},S)\  e^{\ i\varphi'_{\vec{k}} }\Biggr]
,\nonumber \\
{\Pi}_k(\eta)&=& k \left( k^2-V_{\Pi}^{ex}\right)^{-1/4} \Biggl[
S_{ex}^{1/2}\ T\!\cos(S,S^{-1})e^{i\varphi'_{\vec{k}} }-
S_{ex}^{-1/2}\ T\!\sin(S,S^{-1}) e^{i\varphi_{\vec{k}} } \Biggr],
\label{wkbsol4}
\end{eqnarray}
where $\varphi_{\vec{k}}$ and $\varphi'_{\vec{k}}$ now contain an
additional  phase
depending on  the exit time $\eta_{ex}$, and a label {\it ex} means
evaluation at  $\eta =\eta_{ex}$.
The above matching  respects
$S$-duality on the average, in the sense of eq. (\ref{incon}).

%%%%%%%%%%%%%%%%%%%%%%%%%%%%%%%%%%%%%%%%%
\section{Energy spectra}
%%%%%%%%%%%%%%%%%%%%%%%%%%%%%%%%%%%%%%%%%

To complete a calculation of the energy spectrum of the amplified
perturbations  we need to perform a second matching step, to late-time
solutions after horizon re-entry in the decelerated, post-inflationary
era. However, if we are only interested in the leading contribution to
the final energy spectrum,
we can estimate it by using the well-known phenomenon of freezing of
perturbations.
Solutions (\ref{wkbsol4}) always contain a time-independent part
coming from the
first term of the $\ T\!\cos $ expansion (\ref{tsincos3}), and always  
either the constant part of  $\Psi$ or the constant part of $\Pi$
dominates   the Hamiltonian
in the large wavelength limit $ |k \eta| \ll 1$.

In order to illustrate this important point, we  parametrize the
background during the phase of accelerated evolution with a
power-like behaviour of the pumping field, $S^{1/2} \sim |\eta|^\alpha$
for $\eta <0$, $\eta \rightarrow 0_-$. From the general solution
(\ref{wkbsol3}) we obtain, in the limit $|k\eta| \ll 1$,
\begin{equation}
{\widehat\Psi}_k= \widehat
A_k|\eta|^\alpha+ {k|\eta|\over 1-2\alpha } \widehat B_k
|\eta|^{-\alpha} , ~~~~~~~~
{\widehat\Pi}_k= k
\left(\widehat B_k|\eta|^{-\alpha}-
{k|\eta|\over 1+ 2\alpha }\widehat A_k|\eta|^{\alpha}
\right),
\label{311}
\end{equation}
with logarithmic corrections for $\alpha = \pm 1/2$.
Computing   the Hamiltonian density (\ref{313})
we obtain, for $ |k \eta| \ll 1$,
\begin{eqnarray}
H &=&\frac{1}{2} \sum_{\vec{k}} k^2
 \Biggl\{ |\widehat A_k|^2  |\eta|^{2\alpha}\left[1 + O(|k\eta|^2)
\right]
 + |\widehat B_k|^2  |\eta|^{-2\alpha}\left[1 + O(|k\eta|^2) \right]
+  O\left(|\widehat A_k| |\widehat B_k| |k\eta|\right)\Biggr\}
\nonumber  \\
&\simeq & \frac{1}{2} \sum_{\vec{k}} k^2
 \Biggl\{ |\widehat  A_k|^2  |\eta|^{2\alpha} +
|\widehat  B_k|^2  |\eta|^{-2\alpha} \Biggr\}.
\label{314}
\end{eqnarray}
Thus, irrespectively of the value and sign of $\alpha$,
the joint contribution of
the constant modes $\Psi_k \sim \widehat  A_k$ and
$\Pi_k \sim k \widehat B_k$ always provides an
accurate estimate of the leading contribution to the Hamiltonian for
super-horizon wavelengths. Which of the two constant modes
in  (\ref{314}) gives the
dominant contribution varies from case to case. Also, it is not
 always true that the subleading constant mode
 provides the leading correction.

The above statements can be explicitly checked when the pump field
$S$ is given by a power of $\eta$ with different powers
in the different cosmological epochs. In this case, the joining
of solutions at the transition times can be done  analytically,
exact formulae for the Bogoliubov coefficients can be given,
and their duality can be directly checked \cite{8}.
An important point \cite{8} is that, for  backgrounds
with discontinuous derivatives at the transitions,
exact duality is
 satisfied only if continuity is imposed on the
 original perturbation (and on its first derivative) and not
on the rescaled fields  $\widehat\Psi, \widehat\Pi$.

Up to numerical factors, the contributions of
$V_{\Psi}^{ex}$, $V_{\Pi}^{ex}$ in (\ref{wkbsol4})
can be neglected, and one can thus approximate the normalized
vacuum fluctuations outside the horizon by
truncating to $1$ the $\ T\!\cos $ expansion and to $0$ the
$\ T\!\sin $ expansion,
\begin{equation}
{\Psi}_k(\eta) =   \left({k}  S_{ex}\right)^{-1/2}
e^{i\varphi_{\vec{k}}} \, , ~~~~~~~~~~~~~~~
{\Pi}_k(\eta) = \left({k}  S_{ex}\right)^{1/2}
e^{i\varphi'_{\vec{k}} }.
\label{wkbsol5}
\end{equation}
This truncation reflects just a remnant of $S$-duality, the
transformation  $S_{ex} \rightarrow S_{ex}^{-1}$. It is adequate
 for an approximate computation of the energy
 spectrum after re-entry.

At late times $\eta \rightarrow +\infty$, when $V_{\Psi}$, $V_{\Pi}
\rightarrow 0$ in the post-inflationary phase, the perturbation modes
with $k\eta \gg 1$ oscillate freely inside the horizon, according to the
equations
\begin{equation}
{\widehat\Psi}_k{''}+k^2{\widehat\Psi}_k
 = 0 \, , ~~~~~~~~~~~~~~~
{\widehat\Pi}_k{''}+ k^2 {\widehat\Pi}_k = 0,
\label{hrdscndord}
\end{equation}
with plane-wave solutions
\begin{eqnarray}
{\widehat\Psi}_k  &=&
\frac{1}{\sqrt{k}} \left( C_k e^{-i k \eta}+ D_k e^{i k \eta} \right)
 \, , \nonumber\\
{\widehat\Pi}_k &=& {{\widehat\Psi}'}_k -{S'\over 2S} {\widehat\Psi}_k
\simeq  {{\widehat\Psi}'}_k= i\sqrt{k}
\left(- C_k e^{-i k \eta}+ D_k e^{i k \eta} \right).
\label{prdscndord}
\end{eqnarray}
The matching of these solutions to the approximate solutions
(\ref{wkbsol5}), performed at $\eta=\eta_{re} \sim k^{-1}$,  imposes
the conditions
\begin{equation}
{\widehat\Psi}_k(\eta_{re})= {1\over \sqrt k}\left(S_{ex}\over
S_{re}\right)^{-1/2} e^{i\varphi_{\vec{k}} } \, , ~~~~~~~~~~~
{{\widehat\Psi}'}_k(\eta_{re})= {\sqrt k}\left(S_{ex}\over
S_{re}\right)^{1/2} e^{i\varphi'_{\vec{k}} } \, ,
\label{43}
\end{equation}
from which we obtain $C_k, D_k$. The final amplified
vacuum fluctuation amplitude,
in the regime $k\eta \gg1$, is thus given by
\begin{eqnarray}
{\widehat\Psi}_k(\eta)  &=&
\frac{1}{\sqrt{k}} \left[\left(S_{ex}\over S_{re}\right)^{-1/2}
\cos (k\eta)~ e^{i\varphi_{\vec{k}} }  +  \left(S_{ex}\over
S_{re}\right)^{1/2}
\sin(k\eta)~ e^{i\varphi'_{\vec{k}} }  \right]\, , \nonumber\\
{\widehat\Pi}_k(\eta)  &=&
{\sqrt{k}} \left[ \left(S_{ex}\over S_{re}\right)^{1/2}
\cos (k\eta)~ e^{i\varphi'_{\vec{k}} } - \left(S_{ex}\over
S_{re}\right)^{-1/2}
\sin(k\eta)~ e^{i\varphi_{\vec{k}} }\right]
\label{44}
\end{eqnarray}
(we have absorbed into $\varphi_k$ and $\varphi_k'$ an additional
phase factor $k\eta_{re}$, arising from the matching).

The corresponding averaged Hamiltonian density, for a mode $k$,
\begin{equation}
\langle H_k\rangle ={1\over 2}\left(\langle|{\widehat\Pi}_k|^2\rangle
+k^2 \langle  |{\widehat\Psi}_k|^2\rangle \right) = k\left(
{S_{ex}\over S_{re}} + {S_{re}\over S_{ex}}\right),
\label{45}
\end{equation}
is invariant under  $S_{ex}\rightarrow S_{ex}^{-1}$, $S_{re}
\rightarrow S_{re}^{-1}$, and also under an overall rescaling of $S$. The
same invariances thus characterize the spectral energy distribution,
$d\rho_k/d\ln k$ = $(k^3/a^4) \langle H_k\rangle$,
which can be written in terms of the proper energy $\omega =k/a$ as
\begin{equation}
{d\rho(\omega)\over d  \ln \omega} = \omega^4 \left[
{S_{ex}(\omega)\over S_{re}(\omega)} + {S_{re}(\omega)
\over S_{ex}(\omega)}\right]\simeq  \omega^4 ~{\rm Max}
\Biggl\{ {S_{ex}\over S_{re}} , {S_{re}\over S_{ex}}\Biggr\}.
\label{46}
\end{equation}

The approximate expression (\ref{46}) reproduces, up to numerical factors
$O(1)$, the leading contribution to all known particle spectra for
various models of
cosmological evolution. Consider, for instance, graviton
production in a transition at $\eta=\eta_*$ from de Sitter inflation  
to radiation, at constant dilaton.
In that case  $S=a^2$,  $S_{ex}/S_*=(k\eta_*)^2=(k/k_*)^2$ and
$S_*/S_{re}=(k/k_*)^2 $,
where $k_*=\eta_*^{-1}$ is the maximal amplified frequency.
Therefore
$S_{re}/S_{ex}=(k_*/k)^4$.
For all modes
$\omega <\omega_*$, where $\omega_*=k_*/a$, ${\rm Max}
\{ {S_{ex}/ S_{re}} , {S_{re}/ S_{ex}}\}=
S_{re}/S_{ex}=(\omega_*/\omega)^4$, and
we  recover the well-known flat Harrison-Zeldovich spectrum 
\cite{quantum}, 
$d\rho/d\ln \omega \simeq \omega_*^4$.

If, instead, we consider the transition from a phase of
dilaton-driven inflation to
radiation, typical of string cosmology, we can recover from (\ref{46})
the leading contribution to known spectra for photons \cite{2},
gravitons \cite{7}
and axions \cite{3,9,8}. For heterotic photons,
in particular, and more generally for gauge bosons coming from  extra  
dimensions,
$S(\eta) = e^{-\phi (\eta)}$ coincides with the real part of the
usual $S$-field of string
theory, and the leading contribution to the energy spectrum
(\ref{46}) is the same for two
cosmologies that arrive at the same final $S$  from above (weak
coupling) or from below (strong coupling). In the two cases the energy
is predominantly stored in the electric and in the magnetic field,
respectively. From the approximate expression (\ref{46}) we can also
recover the leading contribution to the energy spectrum for more
general classes
of backgrounds and types of perturbations \cite{9,8}.

%%%%%%%%%%%%%%%%%%%%%%%%%%%%%%%%%%%%%%%%%
\section{Lower bounds on  energy spectra}
%%%%%%%%%%%%%%%%%%%%%%%%%%%%%%%%%%%%%%%%%

Since the approximate expression for the energy spectra, eq. (\ref{46}),
contains, as a direct result of the remnant duality
symmetry, a factor of the form
$x+x^{-1}$,  an interesting consequence of the remnant duality
symmetry is the existence of a uniform and {\it model-independent}
lower bound
for the energy spectrum:
\begin{equation}
{d\rho \over d \ln \omega} ~\gaq~ \omega^4,
\label{47}
\end{equation}
valid for all models of backgrounds and all types of perturbations:
 it represents the minimal
spectrum corresponding to the production of one particle per mode, per
polarization and per unit phase-space volume \cite{11}. For some
particular model of background, and for a given type of perturbation,  
eq. (\ref{46}) can  provide a less trivial lower bound,
corresponding to a ``minimal" spectrum flatter than $\omega^4$.
In such cases the  frequency
dependence of the lower bound is, in general, model-dependent.

Consider, in particular, the amplification of tensor perturbations
in a string cosmology background in which the initial dilaton-driven
phase, with $S\sim \eta$, is followed at $\eta=\eta_s$ by a
high-curvature string phase, with final transition to the
radiation-dominated era at $\eta=\eta_1$. During the string phase,
a total redshift $z_s=|a_1/a_s|>1$ is accumulated,
the curvature is approximately constant,
and the dilaton grows by the ratio
$\exp(\phi_1-\phi_s)/2=g_1/g_s>1$. Equation (\ref{46}) gives, in that
case:
\begin{equation}
{d\rho \over d \ln \omega} \simeq \omega^4
{\rm Max}
\Biggl\{ {\omega_1\over \omega} \left(g_s\over g_1 \right)^2 z_s^3
,{\omega\over \omega_1} \left(g_1\over g_s \right)^2 z_s^{-3}\Biggr\}
\; , \; \; \omega < \omega_s \equiv \omega_1/z_s  .
\label{48}
\end{equation}
The assumption that the pump field keeps growing during the string
phase, $z_sg_s/g_1 >1$, then leads  to the well-known cubic ($n=4$)  
graviton spectrum \cite{7}, whose lower bound,
\begin{equation}
{d\rho \over d \ln \omega} ~\gaq~ \omega_1^4 \left(g_s\over g_1
\right)^2 \left({\omega \over \omega_s}\right)^3,
\label{49}
\end{equation}
is more constraining than eq. (\ref{47}). The ``minimal" spectrum is
${d\rho / d \ln \omega} =\omega_1 \omega^3$ and corresponds to the
limiting case in which $g_s \rightarrow g_1$ and $z_s \rightarrow 1$,
namely to a model with almost no intermediate string phase.
Note that, for gravitons, the largest subleading contribution 
is not given 
by the second term in the bracket in eq. (\ref{48}) \cite{7,9} but,
instead, can be extracted by carefully matching  solutions at $\eta =
\eta_1$. When this is done, the more stringent bound can be seen
as a consequence of a
remnant symmetry acting on both $z_s$ and  $g_s/g_1$.

An important point is that the duality-symmetric form
(\ref{46}) of the spectrum, which provides an accurate
order-of-magnitude estimate for the energy spectra produced from
perturbations whose  action is given by (\ref{spertact}), can also
provide a lower bound on the total energy density of
long-wavelength perturbations, even when the background
evolution includes a high-curvature phase in which the action and
perturbation equations are not known.
The argument goes as follows.

The result (\ref{46}) was  obtained by approximating the
evolution of $\Psi$ and $\Pi$ by their constant modes outside the
horizon. The approximation was motivated by the explicit solution  of
the perturbation equations. When the background enters a
high-curvature phase, which requires a  corrected
effective action for an accurate description, the exact form of the
perturbation equations and
their solutions are not always known, or in any case difficult to
obtain in practice. We expect that, for any type of action, the
perturbation equations should always admit constant solutions
whenever  spatial  gradients are
negligible with respect to time derivatives,
 in agreement with the general physical principle of freezing of
perturbations outside the horizon.

For metric perturbations,
a formal argument can be given to justify our conjecture.
Since the action is invariant under gauge (i.e. coordinate)
transformations, a particular set of allowed perturbations are
 infinitesimal gauge transforms
of the background itself. Of course, these are spurious perturbations  
without any physical content. It is possible, however, to
construct physical, large-wavelength perturbations,
which approach the infinitesimal gauge transformations in the
infinite-wavelength limit. By continuity, these  large-wavelength
gravitational waves are approximate solutions of the perturbation
equations.
The amount by which they deviate from exact solutions away from  the
limit is controlled by the only dimensionless parameter present, i.e. by
$k \eta$.  We may thus conclude that, at least for metric perturbations,
the constant mode in $\Psi$ is always present for perturbations
sufficiently outside the horizon. Canonical Poisson brackets then
require that the same  be true for $\Pi$. Although we cannot exclude
that other modes in $\Psi$ and $\Pi$ will dominate over the constant
modes, assuming that this does not happen will provide a lower bound
on the amplified  spectra in the presence of  a high-curvature phase.

For tensor perturbations, this argument has been confirmed within an
explicit string theory model of quadratic curvature corrections
\cite{12}. The results of a numerical integration show that $\Psi$
becomes frozen outside the horizon even after including the
higher-curvature terms in the perturbation equations.
The final normalized amplitude is larger,
by a numerical factor, than the estimate (\ref{wkbsol5}), so that the  
spectrum
(\ref{46}) indeed
represents a lower bound for the total energy density.

%%%%%%%%%%%%%%%%%%%%%%%%%%%%%%%%%%%%%%%%%
\section{Conclusion}
%%%%%%%%%%%%%%%%%%%%%%%%%%%%%%%%%%%%%%%%

We have shown that the equations of
cosmological perturbation theory, to lowest order in the strength of
perturbations and in their derivatives, are  invariant
under a duality transformation acting on the classical moduli of the
background and interchanging the perturbation with its conjugate
momentum.  As a consequence of this invariance, the large-wavelength  
part of the Hamiltonian is always dominated by the
sum of the  contributions of the frozen modes of the perturbation  and
of its conjugate momentum.

The energy spectrum obtained by truncating the
solutions of the perturbation equations to the constant modes is
characterized by a
residual duality symmetry; it reproduces known results of produced
particle spectra and can provide a useful lower bound on particle
production, when our knowledge of the detailed dynamical history of
the background is approximate or incomplete.

\vskip 2 cm

\section*{Acknowledgements}

We are grateful to A. Buonanno, K.A. Meissner and C. Ungarelli for
interesting discussions. R. B. is supported in part by the  Israel
Science Foundation administered by the Israel Academy of Sciences and
Humanities.

\end{document}